\documentstyle[12pt]{article}
\begin{document}
\begin{flushright}
hep-th/0402123\\
SNB/February/2004
\end{flushright}
\vskip 2.5cm
\begin{center}
{\bf \Large {Nilpotent Symmetries For A Free Relativistic 
Particle\\ In Augmented Superfield Formalism}}

\vskip 2.5cm

{\bf R.P.Malik}
\footnote{ E-mail address: malik@boson.bose.res.in  }\\
{\it S. N. Bose National Centre for Basic Sciences,} \\
{\it Block-JD, Sector-III, Salt Lake, Calcutta-700 098, India} \\

\vskip 2cm

\end{center}

\noindent
{\bf Abstract}: 
In the framework of the augmented superfield formalism, 
the local, covariant, continuous and off-shell (as well as 
on-shell) nilpotent (anti-)BRST symmetry transformations are derived 
for a  $(0 + 1)$-dimensional free scalar relativistic particle 
that provides a prototype physical example for the 
more general reparametrization invariant
string- and gravitational theories. The trajectory (i.e. the world-line) of the 
free particle, parametrized by a monotonically increasing evolution parameter 
$\tau$, is embedded in a $D$-dimensional flat Minkowski target manifold. This 
one-dimensional system is considered on 
a $(1 + 2)$-dimensional supermanifold parametrized by an even element 
$\tau$ and a couple of odd elements ($\theta$ and $\bar\theta$) of a 
Grassmannian algebra. The horizontality condition and the invariance of
the conserved (super)charges on the (super)manifolds play very crucial
roles in the above derivations of the nilpotent symmetries. The geometrical 
interpretations for the nilpotent (anti-)BRST charges are
provided in the framework of augmented superfield approach.\\

\baselineskip=16pt

\vskip .7cm



\newpage

\noindent
{\bf 1 Introduction}\\

\noindent
The principle of local gauge invariance, present at the heart of
all {\it interacting} 1-form gauge theories, provides a precise theoretical 
description of the three (out of four) fundamental interactions of nature.
The {\it crucial} interaction term for these interacting theories arises 
due to the coupling of the 1-form gauge fields to the conserved matter 
currents so that the local gauge invariance could be maintained.
In other words, the requirement of the local gauge invariance 
(which is more general than its global counterpart)
enforces a theory to possess an interaction term (see, e.g., [1]).
One of the most attractive approaches to 
covariantly quantize such kind of gauge theories is the 
Becchi-Rouet-Stora-Tyutin (BRST) formalism where the unitarity and ``quantum''
gauge (i.e. BRST) invariance are respected together at any arbitrary order of 
the perturbation theory (see, e.g., [2]). The BRST formalism is 
indispensable in the context of modern developments in topological field
theories, topological string theories, supersymmetric gauge theories, 
reparametrization invariant theories 
(that include D-branes and M-theories), etc., (see, e.g., [3-5] and references
therein for details).

We shall be concentrating, in our present investigation,  only on the 
geometrical aspects of the BRST formalism in the framework of augmented 
superfield formulation because such a study is expected
to shed some light on the abstract mathematical structures 
behind the BRST formalism in a more intuitive and illuminating
fashion. The usual superfield approach [6-13] to  
the BRST scheme provides the geometrical 
origin and interpretation for the conserved 
($\dot Q_{(a)b} = 0$), nilpotent ($Q_{(a)b}^2 = 0$) and anticommuting
($Q_b Q_{ab} + Q_{ab} Q_b = 0$) (anti-)BRST charges $Q_{(a)b}$
which generate local, covariant, continuous, nilpotent ($s_{(a)b}^2 = 0$)
and anticommuting ($s_b s_{ab} + s_{ab} s_b = 0$)
(anti-)BRST transformations $s_{(a)b}$ for {\it only} the gauge field
and (anti-)ghost fields of an interacting gauge theory. This is achieved
by exploiting the so-called horizontality condition [6-13] (which has been
christened as the ``soul-flatness''  condition in [14]). In fact, these 
attempts (see, e.g., [6-14]) have been primarily made to gain an insight 
into the existence of a possible connection between the ideas behind the
supersymmetries and the (anti-)BRST symmetries. As a bonus and by-product,
one obtains the nilpotent (anti-)BRST symmetry transformations for the gauge- 
and the (anti-)ghost fields for the (anti-)BRST invariant Lagrangian density of
a given gauge theory. The horizontality condition requires the
$(p + 1)$-form super-curvature, defined on the $(D + 2)$-dimensional
supermanifold, {\it to be equal} to the $(p + 1)$-form ordinary curvature
for a $p$-form ($ p = 1, 2, 3....$)
gauge theory, defined on an ordinary $D$-dimensional spacetime manifold.
In the above, the $(D + 2)$-dimensional supermanifold is parametrized by the
$D$-number of spacetime {\it even} coordinates and two Grassmannian
{\it odd} coordinates.
Basically, the hoizontality condition owes its origin to the (super)exterior
derivatives $(\tilde d) d$ (with $\tilde d^2 = 0, d^2 = 0$) which are one of
the three (super) de Rham cohomological operators. In a set of papers
[15-19], all the three (super)cohomological operators have been exploited
to derive the (anti-)BRST symmetries, (anti-)co-BRST symmetries and a bosonic
symmetry for the two-dimensional free Abelian gauge theory in the superfield 
formulation where the generalized versions of the horizontality condition
have been exploited. All the above attempts [6-19], however, 
have  {\it not} yet been able to shed
any light on the nilpotent symmetries that exist for the {\it matter} fields of 
an {\it interacting} gauge theory. Thus, the results of the above approaches
[6-19] are still partial as far as the derivation of
{\it all} the symmetry transformations are concerned.

Recently, in a set of papers [20-22], the restriction due to
the horizontality condition has been augmented with the requirement of the
invariance of matter (super)currents on the (super)manifolds
\footnote{We christen this extended version of the usual
formulation  as the  augmented superfield formalism.}.
The latter
restriction produces the nilpotent (anti-)BRST symmetry transformations for 
the matter fields of a given interacting gauge theory. The salient features
of these requirements are (i) there is a beautiful consistency and
complementarity between the nilpotent
transformations generated by the horizontality
restriction and the requirement of conserved matter (super)currents on the
(super)manifolds. (ii) The geometrical interpretations for the (anti-)BRST
charges $Q_{(a)b}$ as the translation generators
$(\mbox {Lim}_{\bar\theta \to 0} (\partial/\partial\theta))
\mbox {Lim}_{\theta \to 0} (\partial/\partial\bar\theta)$ along the
$(\theta)\bar\theta$-directions of the $(D + 2)$-dimensional supermanifold
remain intact. (iii) The nilpotency of the (anti-)BRST charges is encoded
in a couple of successive translations 
(i.e. $(\partial/\partial\theta)^2 = (\partial/\partial\bar\theta)^2 = 0$)
along either of the two Grassmannian
directions of the supermanifold in the framework of both the restrictions. 
(iv) The anticommutativity of the (anti-)BRST charges (and the transformations
they generate) is captured in the relationship
$(\partial/\partial\theta) (\partial/\partial\bar\theta) +
(\partial/\partial\bar\theta) (\partial/\partial\theta) = 0$ for the validity
of both the restrictions. Thus, it is clear that both the above restrictions
enable us to obtain all the nilpotent symmetry transformations for {\it all}
the fields of an {\it interacting} gauge theory.

The purpose of the present paper is to derive the nilpotent (anti-)BRST 
transformations for {\it all} the fields present in the description of
a free scalar relativistic particle (moving on a world-line) in the 
framework of  augmented superfield formulation [20-22]. This study
is essential primarily
on three counts. First, to check the mutual consistency and
complementarity between (i) the horizontality condition, and (ii) the
invariance of the conserved (super)charges on the (super)manifolds
for this physical system. These were found to be true in the case of
interacting (non-)Abelian gauge theories in two $(1 + 1)$-dimensions (2D)
and four $(3 + 1)$-dimensions (4D) of spacetime [20-22]. Of course, the
latter theories were considered on the four $(2 + 2)$-dimensional-
and six $(4 + 2)$-dimensional supermanifolds, respectively. Second, to 
generalize our earlier
works [20-22] (which were connected {\it only} with the gauge symmetries) to
the case where the gauge symmetries as well as the reparametrization symmetries
co-exist in the theory. Finally, to tap the potential and power
of the above restrictions for the case of a new physical system 
defined on a new three $(1 + 2)$-dimensional supermanifold which is
somewhat different from our earlier considerations of the interacting
(non-)Abelian gauge theories on four $(2 + 2)$-
and six $(4 + 2)$-dimensional supermanifolds.

The contents of our present paper are organized as follows. To set up the
notations and conventions, we recapitulate the bare essentials of the
(anti-)BRST symmetries for the free relativistic particle in Section 2. This 
is followed, in Section 3,
by the derivations of the (anti-)BRST symmetries for the gauge (einbein)-
and (anti-)ghost fields in the framework of superfield formulation.
Section 4 is devoted to the derivation of the 
above nilpotent symmetry transformations for the target space variables.
Finally, we make some concluding remarks in Section 5 and point out a few
future directions for our further investigations.\\

\noindent
{\bf 2 Nilpotent (anti-)BRST symmetries in Lagrangian formulation}\\

\noindent
Let us begin with the various equivalent forms of the 
gauge- and reparametrization
invariant Lagrangians for a free scalar relativistic particle moving 
on a world-line that is embedded in a $D$-dimensional
flat Minkowski target manifold.  These specific Lagrangians are [23,24]
$$
\begin{array}{lcl}
L_{0} = m \; (\dot x^2)^{1/2}, \qquad
L_{f} = p_\mu \dot x^\mu - \frac{1}{2}\;e\; (p^2 - m^2), \qquad
L_{s} = \frac{1}{2}\; e^{-1}\; (\dot x)^2 + \frac{1}{2}\; e \; m^2.
\end{array} \eqno(2.1)
$$
In the above, the mass-shell condition ($p^2 - m^2 = 0$) and the force free
(i.e. $\dot p_\mu = 0$) motion of the {\it free} relativistic particle
are a couple  of  common 
features for (i) the Lagrangian with the square root $L_{0}$, (ii) the
first-order Lagrangian density $L_f$, and (iii)
the second-order Lagrangian $L_s$. Except for the  mass (i.e. the analogue of 
the cosmological constant) parameter $m$, the target space canonically 
conjugate coordinates $x^\mu (\tau)$
(with $\mu = 0, 1, 2......D-1$) as well as the momenta $p_\mu (\tau)$ and
the einbein field $e (\tau)$ are the functions of the monotonically increasing
parameter $\tau$ that characterizes the trajectory (i.e. the world-line)
of the free scalar 
relativistic particle. Here $\dot x^\mu = (d x^\mu/d \tau)$ are the
generalized versions of the
``velocity'' of the particle. All the above 
dynamical variables as well as the mass parameter $m$ are the {\it even}
elements of the Grassmann algebra. The first- and the second-order Lagrangians 
are  endowed with the first-class constraints $\Pi_e \approx 0$
and $p^2 - m^2 \approx 0$ in the language of Dirac's classification 
scheme where $\Pi_e$ is the canonical conjugate
momentum corresponding to the einbein field $e(\tau)$. The existence of
the first-class constraints on this physical system
establishes the fact that this reparametrization
invariant theory of the free relativistic particle is a {\it gauge} theory
\footnote{It can be readily seen that under the infinitesimal transformations
$\delta_r x_\mu = \epsilon \dot x_\mu, 
\delta_r p_\mu = \epsilon \dot p_\mu, 
\delta_r e = (d/d\tau) [(\epsilon e)]$ generated by the 
basic reparametrization $\tau \to
\tau - \epsilon (\tau)$, the Lagrangian density $L_f$ transforms to
$\delta_r L_f = (d/d \tau) [( \epsilon L_f)]$. Similarly, under the gauge
transformations $\delta_g x_\mu = \xi p_\mu, \delta_g p_\mu = 0, \delta_g e
= \dot \xi $, the Lagrangian density $L_f$ transforms to a total derivative.
Both these transformations are equivalent (with the identification
$\xi = e \epsilon$) for the free (i.e. $\dot p_\mu = 0$) 
relativistic particle because
both the above transformations owe their origin to the mass-shall condition
$p^2 - m^2 = 0$. It is evident that $\dot p_\mu =0$ and $p^2 -m^2 = 0$ are
a couple of salient features in the physical description
of a free relativistic particle.}.
For the covariant canonical quantization of such systems, 
one of the most elegant
and suitable approaches is the Becchi-Rouet-Stora-Tyutin
(BRST) formalism. The (anti-)BRST invariant
Lagrangian $L_B$ corresponding to the above first-order Lagrangian
$L_f$ is as follows (see, e.g., [24]  and references therein)
$$
\begin{array}{lcl}
L_{B} =   p_\mu \dot x^\mu - \frac{1}{2}\;e\; (p^2 - m^2) + b\; \dot e
+ \frac{1}{2}\; b^2 - i\; \dot {\bar c}\; \dot c,
\end{array} \eqno(2.2)
$$
where the {\it even} element $b(\tau)$ is the Nakanishi-Lautrup auxiliary
field and the (anti-)ghost fields $(\bar c)c$ are the {\it odd} elements of the
Grassmann algebra (i.e. $\bar c^2 = c^2 = 0, c \bar c + \bar c c = 0$). 
The above Lagrangian density (2.2) respects the following off-shell nilpotent
$(s_{(a)b}^2 = 0$) (anti-)BRST $s_{(a)b}$ symmetry transformations
\footnote{We follow here the notations and conventions adopted by
Weinberg [25]. In its totality, the true nilpotent 
(anti-)BRST transformations $\delta_{(A)B}$ are the product of an 
(anticommuting) spacetime
independent parameter $\eta$ and the nilpotent transformations $s_{(a)b}$.
It is clear that $\eta$ commutes with all the bosonic (even) fields of the
theory and anti-commutes with fermionic (odd) fields
(i.e. $\eta c + c \eta = 0, \eta \bar c + \bar c \eta = 0$).}
(with $s_b s_{ab} + s_{ab} s_b = 0$) (see, e.g., [24])
$$
\begin{array}{lcl}
s_b x_\mu &=& c p_\mu, \qquad s_b c = 0, \qquad s_b p_\mu = 0, \qquad
s_b \bar c = i b, \qquad s_b b = 0, \qquad s_b e = \dot c, \nonumber\\
s_{ab} x_\mu &=& \bar c p_\mu, \qquad s_{ab} \bar c = 0, \quad 
s_{ab} p_\mu = 0, \quad
s_{ab} c = - i b, \qquad s_{ab} b = 0, \;\quad s_{ab} e = \dot {\bar c},
\end{array} \eqno(2.3)
$$
because the Lagrangian density (2.2) transforms to 
the following total derivatives:
$$
\begin{array}{lcl}
s_b L_B = {\displaystyle \frac{d} {d \tau}} \;
\Bigl [ \frac{1}{2}\; c\; (p^2 + m^2) + b\; \dot c \Bigr ], \qquad\;
s_{ab} L_B = {\displaystyle \frac{d} {d \tau}} \;
\Bigl [ \frac{1}{2}\; \bar c\; (p^2 + m^2) + b\; \dot {\bar c} \Bigr ].
\end{array} \eqno(2.4)
$$
The on-shell ($\ddot c = 0, \ddot {\bar c} = 0$) nilpotent 
($\tilde s_{(a)b}^2 = 0$) (anti-)BRST symmetry
transformations $\tilde s_{(a)b}$ can be derived from (2.3) by the substitution
$b = - \dot e$ as  listed below
$$
\begin{array}{lcl}
\tilde s_b x_\mu &=& c p_\mu, \qquad \tilde s_b c = 0, \;\qquad 
\tilde s_b p_\mu = 0, \;\qquad
\tilde s_b \bar c = - i \dot e, \qquad \tilde s_b e = \dot c, \nonumber\\
\tilde s_{ab} x_\mu &=& \bar c p_\mu,
\qquad \tilde s_{ab} \bar c = 0, \qquad 
\tilde s_{ab} p_\mu = 0, \qquad
\tilde s_{ab} c = + i \dot e, \qquad \tilde s_{ab} e = \dot {\bar c},
\end{array} \eqno(2.5)
$$
which turn out to be the symmetry transformations for the
following Lagrangian density
$$
\begin{array}{lcl}
\tilde L_{B} =   p_\mu \dot x^\mu - \frac{1}{2}\;e\; (p^2 - m^2) 
- \frac{1}{2}\; (\dot e)^2 - i\; \dot {\bar c}\; \dot c.
\end{array} \eqno(2.6)
$$
It should be noted that, for the on-shell nilpotent version of the (anti-)BRST
transformations (2.5) and corresponding Lagrangian density (2.6), it is only
the $ b (= - \dot e)$ field that has been replaced. Rest of the transformations
of (2.3) remain intact. In the BRST quantization 
procedure, the first-class constraints $\Pi_e = b \approx 0$ as well
as $p^2 - m^2 = - 2 \dot b \approx 0$ appear as the constraints on the physical
states when one requires that the conserved and off-shell nilpotent
BRST charge $Q_b = \frac{1}{2}\; c\; (p^2 - m^2) 
+ b\; \dot c \equiv b \dot c - \dot b c$ should annihilate 
the physically meaningful states in the quantum Hilbert space of states.
The conservation of the off-shell
nilpotent ($Q_b^2 = 0$) BRST charge $Q_b$ on any arbitrary {\it unconstrained}
manifold is ensured by exploiting the equations of motion
$\dot p_\mu = 0, \dot b = - \frac{1}{2} (p^2 - m^2), 
\ddot c =  \ddot {\bar c} = 0, b + \dot e = 0,  \dot x_\mu = e \; p_\mu$.
The expression for the on-shell nilpotent BRST charge $\tilde Q_b$
can be obtained from $Q_b$ by the substitution $ b = - \dot e$. Both
the off-shell as well as the on-shell nilpotent symmetry transformations can
be succinctly expressed in terms of the conserved ($\dot Q_r 
= \dot {\tilde Q}_r = 0$) charges ($Q_r, \tilde Q_r$)
and the generic field $\Psi = x_\mu, p_\mu, c, \bar c, b, e$, as
$$
\begin{array}{lcl}
s_r \Psi = - i\; [\Psi, Q_r]_{\pm}, \qquad
\tilde s_r \Psi = - i\; [\Psi, \tilde Q_r]_{\pm}, \qquad 
r = b, ab,
\end{array} \eqno(2.7)
$$
where $Q_{ab}$ and $\tilde Q_{ab}$ are the off-shell- and the 
on-shell nilpotent versions of the
anti-BRST charges which can be obtained from their counterpart BRST charges
by the substitutions: $c \to \bar c, \bar c \to c$. The explicit expressions
for the conserved charges $Q_{(a)b}$ and $\tilde Q_{(a)b}$ are not
required for our present discussions but they can be found in [24].
The $\pm$ signs, 
present as the subscripts on the above square
brackets, stand for the (anti-)commutators for the generic field $\Psi$
being (fermionic)bosonic in nature.\\

\noindent
{\bf 3 Symmetries for the gauge- and 
(anti-)ghost fields in superfield approach}\\

\noindent
We begin here with a general three ($1 + 2$)-dimensional supermanifold
parametrized by the superspace coordinates $Z = (\tau, \theta, \bar \theta)$
where $\tau$ is an
even (bosonic) coordinate
and $\theta$ and $\bar \theta$ are the two odd (Grassmannian) coordinates
(with $\theta^2 = \bar \theta^2 = 0, 
\theta \bar \theta + \bar \theta \theta = 0)$. On this supermanifold, one can
define a 1-form supervector superfield $\tilde V = d Z (\tilde A)$ with
$\tilde A (\tau,\theta,\bar\theta) = (E (\tau, \theta, \bar \theta), 
\;\Phi (\tau, \theta, \bar \theta), \;\bar \Phi (\tau, \theta, \bar \theta))$
as the component multiplet superfields. The superfields $E, \Phi, \bar \Phi $ 
can be expanded in terms of the basic fields ($e, c, \bar c$) and auxiliary 
field ($b$) along with  some extra secondary fields 
(i.e. $f, \bar f, B, {\cal B}, \bar {\cal B}, s, \bar s, \bar b$), as
(see, e.g., [9])
$$
\begin{array}{lcl}
E\; (\tau, \theta, \bar \theta) &=& e (\tau) 
\;+\; \theta\; \bar f  (\tau) \;+\; \bar \theta\; f (\tau) 
\;+\; i \;\theta \;\bar \theta \;B (\tau), \nonumber\\
\Phi\; (\tau, \theta, \bar \theta) &=& c (\tau) 
\;+\; i\; \theta\; \bar b (\tau)
\;+\; i \;\bar \theta\; {\cal B} (\tau) 
\;+\; i\; \theta\; \bar \theta \;s (\tau), \nonumber\\
\bar \Phi\; (\tau, \theta, \bar \theta) &=& \bar c (\tau) 
\;+\; i \;\theta\;\bar {\cal B} (\tau)\; +\; i\; \bar \theta \;b (\tau) 
\;+\; i \;\theta \;\bar \theta \;\bar s (\tau).
\end{array} \eqno(3.1)
$$
It is straightforward to note that the local 
fields $ f (\tau), \bar f (\tau),
c (\tau), \bar c (\tau), s (\tau), \bar s (\tau)$ on the r.h.s.
are fermionic (anti-commuting) 
in nature and the bosonic (commuting) local fields in (3.1)
are: $e (\tau), B (\tau), {\cal B} (\tau), \bar {\cal B} (\tau),
b (\tau), \bar B (\tau)$. It is unequivocally clear
that, in the above expansion, the bosonic-
 and fermionic degrees of freedom match. This requirement is essential
for the validity and sanctity of any arbitrary supersymmetric theory in the 
superfield formulation. In fact, all the secondary fields will be expressed 
in terms of basic fields due to the restrictions emerging from the application 
of the horizontality condition [9,8]
$$
\begin{array}{lcl} 
\tilde d\; \tilde V = d\; A  = 0, \qquad d = d \tau\; \partial_\tau, \qquad
A = d \tau\; e(\tau), \qquad d^2 = 0,
\end{array} \eqno(3.2)
$$
where the super exterior derivative $\tilde d$ and 
the super connection  1-form $\tilde V$ are defined as
$$
\begin{array}{lcl}
\tilde d &=&  d \tau\; \partial_\tau\;
+ \;d \theta \;\partial_{\theta}\; + \;d \bar \theta \;\partial_{\bar \theta},
\nonumber\\
\tilde V &=& d \tau \;E (\tau , \theta, \bar \theta)
+ d \theta\; \bar \Phi (\tau, \theta, \bar \theta) + d \bar \theta\;
\Phi (\tau, \theta, \bar \theta).
\end{array}\eqno(3.3)
$$
We expand $\tilde d \;\tilde V$, present in the l.h.s. of (3.2), as
\footnote{We have exploited here: $d \tau \wedge d \tau = 0, 
d \tau \wedge d \theta = - d \theta \wedge d \tau, 
d \tau \wedge d \bar \theta = - d \bar\theta \wedge d \tau,
d \theta \wedge d \bar \theta
= + d \bar\theta \wedge d \theta$, etc. together with
 $\partial_\theta^2 = \partial_{\bar\theta}^2 = 0, 
\partial_\theta \partial_{\bar\theta}
+ \partial_{\bar\theta} \partial_\theta = 0$ and
$(d\theta\; \partial_\theta)
\wedge (d \bar\theta \;\partial_{\bar\theta}) = - (d \theta \wedge
d \bar\theta) (\partial_\theta\; \partial_{\bar\theta}) $, etc.} 
$$
\begin{array}{lcl}
\tilde d \;\tilde V  &=& 
(d \tau \wedge d \theta) (\partial_{\tau} \bar \Phi - \partial_{\theta} E) 
- (d \theta \wedge d \theta)\; (\partial_{\theta}
\bar \Phi) + (d \tau \wedge d \bar \theta)
(\partial_{\tau} \Phi - \partial_{\bar \theta} E) \nonumber\\
&-& (d \theta \wedge d \bar \theta) (\partial_{\theta} \Phi 
+ \partial_{\bar \theta} \bar \Phi) 
- (d \bar \theta \wedge d \bar \theta)
(\partial_{\bar \theta} \Phi). 
\end{array}\eqno(3.4)
$$
Ultimately, the application of the horizontality condition
\footnote{It should be noted that the gauge (einbein) field $e (\tau)$ is a
scalar potential depending only on a single parameter $\tau$. This is why the 
curvature $d A = 0$ (because $d\tau \wedge d \tau = 0$). For the 1-form 
Abelian gauge theory where the gauge field is a vector potential $A_\mu (x)$
(defined through $A = dx^\mu A_\mu$),
the 2-form curvature $d A = \frac{1}{2} (dx^\mu \wedge dx^\nu)\; F_{\mu\nu}$
is not equal to zero and
defines the field strength tensor $F_{\mu\nu} = \partial_\mu A_\nu 
- \partial_\nu A_\mu$.}      
($\tilde d \tilde V = d A = 0$) yields
$$
\begin{array}{lcl}
f \;(\tau) &=& \partial_{\tau}\; c \;(\tau) \equiv \dot c (\tau), \qquad 
\bar f\; (\tau) = \partial_{\tau}\;
\bar c\; (\tau) \equiv \dot {\bar c} (\tau),
\qquad \;s\; (x) = \bar s\; (x) = 0,
\nonumber\\
B\; (\tau) &=& \partial_{\tau} b\; (\tau) \equiv \dot b (\tau),
\qquad\;\;\;
b\; (\tau) + \bar b \;(\tau) = 0, \qquad \;\;\;
{\cal B}\; (\tau)  = \bar {\cal B} (\tau) = 0.
\end{array} \eqno(3.5)
$$
The insertion of all the above values in the expansion (3.1) leads to
the derivation of the (anti-)BRST symmetry transformations  for the 
gauge- and (anti-)ghost fields of the theory. This 
statement can be expressed, in an explicit form, as given below
$$
\begin{array}{lcl}
E\; (\tau, \theta, \bar \theta) &=& e (\tau) 
+ \theta\; \dot {\bar c }(\tau) + \bar \theta\;  \dot c (\tau)) 
+ i \;\theta \;\bar \theta \;\dot b (\tau), \nonumber\\
\Phi\; (\tau, \theta, \bar \theta) &=& c (\tau) 
- i\; \theta\;  b (\tau), \nonumber\\
\bar \Phi\; (\tau, \theta, \bar \theta) &=& \bar c (\tau) 
+ i\; \bar \theta \;b (\tau).
\end{array} \eqno(3.6)
$$
In addition, this exercise provides  the physical interpretation for the
(anti-)BRST charges $Q_{(a)b}$ 
as the generators (cf.(2.7)) of translations 
(i.e. $ \mbox{Lim}_{\bar\theta \rightarrow 0} (\partial/\partial \theta),
 \mbox{Lim}_{\theta \rightarrow 0} (\partial/\partial \bar\theta)$)
along the Grassmannian
directions of the supermanifold. Both these observations can be succinctly 
expressed, in a combined way, by re-writing the super expansion (3.1) as
$$
\begin{array}{lcl}
E\; (\tau, \theta, \bar \theta) &=& e (\tau) 
+ \;\theta\; (s_{ab} e(\tau)) 
+ \;\bar \theta\; (s_{b} e(\tau)) 
+ \;\theta \;\bar \theta \;(s_{b} s_{ab} e (\tau)), \nonumber\\
\Phi\; (\tau, \theta, \bar \theta) &=& c (\tau) \;
+ \; \theta\; (s_{ab} c (\tau))
\;+ \;\bar \theta\; (s_{b} c (\tau)) 
\;+ \;\theta \;\bar \theta \;(s_{b}\; s_{ab}  (\tau)), 
\nonumber\\
\bar \Phi\; (\tau, \theta, \bar \theta) &=& \bar c (\tau) 
\;+ \;\theta\;(s_{ab} \bar c (\tau)) \;+\bar \theta\; (s_{b} \bar c (\tau))
\;+\;\theta\;\bar \theta \;(s_{b} \;s_{ab} \bar c (\tau)).
\end{array} \eqno(3.7)
$$
It should be noted that the third and fourth terms of the expansion
for $\Phi$ and the second and fourth terms of the expansion for $\bar \Phi$
are zero because ($s_b c = 0, \; s_{ab} \bar c = 0$).

Let us now focus on the derivation of the on-shell nilpotent
(anti-)BRST symmetry transformations $\tilde s_{(a)b}$ of (2.5) in the
framework of superfield approach. To this end in mind, we begin with the 
chiral limit $ \theta \to 0$ of the expansions (3.1) and 
definition (3.3),  as 
$$
\begin{array}{lcl}
&& (E)|_{(c)}\; (\tau, \bar \theta) = e (\tau) 
+ \bar \theta\; f (\tau), \qquad\;\;\;
(\Phi)|_{(c)}\; (\tau, \bar \theta) = c (\tau) 
+ i \;\bar \theta\; {\cal B} (\tau), \nonumber\\
&&(\bar \Phi)|_{(c)}\; (\tau, \bar \theta) = \bar c (\tau) 
+ i\; \bar \theta \;b (\tau), \qquad \;\;\;\
(\tilde d)|_{(c)} = d\tau\; \partial_\tau +
d \bar\theta\; \partial_{\bar\theta}, \nonumber\\
&&(\tilde V)|_{(c)} (\tau, \bar\theta)
= d \tau\; (E)|_{(c)} \;(\tau, \bar\theta)
 + \;d\bar\theta\; (\Phi)|_{(c)}\; (\tau, \bar\theta).
\end{array} \eqno(3.8)
$$
The horizontality condition $(\tilde d)|_{(c)} (\tilde V)|_{(c)} = d A = 0$,
leads to the following conditions
$$
\begin{array}{lcl}
\partial_\tau (\Phi)|_{(c)} = \partial_{\bar\theta} (E)|_{(c)}
\Rightarrow f (\tau) = \dot c (\tau), \qquad
\partial_{\bar\theta} (\Phi)|_{(c)} = 0 \Rightarrow {\cal B} (\tau) = 0,
\end{array} \eqno(3.9)
$$
which emerge from the explicit expression for 
$ (\tilde d) |_{(c)} (\tilde V) |_{(c)} = 0$ as given below
$$
\begin{array}{lcl}
(\tilde d)|_{(c)} (\tilde V)|_{(c)} = 
(d \tau \wedge d \bar \theta)\;
\bigl [\;\partial_{\tau} (\Phi)|_{(c)} - \partial_{\bar \theta} (E)|_{(c)}
\;\bigr ] 
- (d \bar \theta \wedge d \bar \theta)
\; \bigl [\;\partial_{\bar \theta} (\Phi)|_{(c)} \;\bigr ] \equiv 0.
\end{array}\eqno(3.10)
$$
It is clear that two of the three extra fields 
present in the expansion (3.8)
are found in (3.9) due to the horizontality restriction. However, the extra
field $b(\tau)$ is {\it not} determined by the above
 condition. Fortunately, the 
equation of motion $b (\tau) + \dot e (\tau) = 0$, emerging from the Lagrangian
density (2.2), comes to our rescue. With these inputs,
we obtain the following form of the expansion (3.1) {\it vis-{\` a}-vis}
the on-shell nilpotent transformations (2.5):
$$
\begin{array}{lcl}
&& (E)|_{(c)}\; (\tau, \bar \theta) = e (\tau) 
+ \bar \theta\; (\tilde s_b e(\tau)), \qquad\;\;\;
(\Phi)|_{(c)}\; (\tau, \bar \theta) = c (\tau) 
+ \;\bar \theta\; (\tilde s_b c (\tau)), \nonumber\\
&&(\bar \Phi)|_{(c)}\; (\tau, \bar \theta) = \bar c (\tau) 
+ \; \bar \theta \;(\tilde s_b c(\tau)). 
\end{array} \eqno(3.11)
$$
The above expansion, together with the help of (2.7), provides the
geometrical interpretation for the conserved and
on-shell nilpotent BRST charge $\tilde Q_{b}$ as the generator of translation
(i.e. $(\partial/\partial \bar\theta)$) for the {\it chiral} superfields 
(3.8) along the $\bar\theta$-direction of the
two $(1 + 1)$-dimensional chiral super sub-manifold, parametrized by an even
variable $\tau$ and an odd variable $\bar\theta$. In a similar fashion,
one can derive the on-shell nilpotent anti-BRST transformations of
(2.5) by exploiting the anti-chiral limit of (3.1). In fact, in the
limit $\bar\theta \to 0$, the expansions in (3.1) and definition in (3.3)
lead to
$$
\begin{array}{lcl}
&& (E)|_{(ac)}\; (\tau,  \theta) = e (\tau) 
+ \theta\; \bar f (\tau), \qquad\;\;\;
(\Phi)|_{(ac)}\; (\tau,  \theta) = c (\tau) 
- i \; \theta\; b  (\tau), \nonumber\\
&&(\bar \Phi)|_{(ac)}\; (\tau, \theta) = \bar c (\tau) 
+ i\; \theta \;\bar {\cal B} (\tau), \qquad \;\;\;\
(\tilde d)|_{(ac)} = d\tau\; \partial_\tau +
d \theta\; \partial_{\theta}, \nonumber\\
&&(\tilde V)|_{(ac)} (\tau, \theta)
= d \tau\; (E)|_{(ac)} \;(\tau, \theta)
 + \;d \theta\; (\bar \Phi)|_{(ac)} \;(\tau, \theta).
\end{array} \eqno(3.12)
$$
It should be noted that, from our earlier consideration, it was found  that
$b (\tau) + \bar b(\tau) = 0$. This relation has been exploited here to replace
$\bar b(\tau)$ in the expansion of $(\Phi)|_{(ac)}$ by $- b(\tau)$.
The horizontality condition $(\tilde d)|_{(ac)} (\tilde V)|_{(ac)} = d A = 0$,
leads to the following conditions
$$
\begin{array}{lcl}
\partial_\tau (\bar \Phi)|_{(ac)} = \partial_{\theta} (E)|_{(ac)}
\Rightarrow \bar f (\tau) = \dot {\bar c} (\tau), \qquad
\partial_{\theta} (\bar \Phi)|_{(ac)} = 0 \Rightarrow \bar {\cal B} (\tau) = 0,
\end{array} \eqno(3.13)
$$
which emerge from the explicit expression for 
$(\tilde d)|_{(ac)} (\tilde V)|_{(ac)} = 0$ as listed below
$$
\begin{array}{lcl}
(\tilde d)|_{(ac)} (\tilde V)|_{(ac)} = 
(d \tau \wedge d \theta)\;
\bigl [\;\partial_{\tau} \bar\Phi)|_{(ac)} 
- \partial_{ \theta} (E)|_{(ac)}\; \bigr ]
- (d \theta \wedge d  \theta)\;
\bigl [\;\partial_{ \theta} (\bar \Phi)|_{(ac)}\; \bigr ] \equiv  0.
\end{array}\eqno(3.14)
$$
It can be seen that two of the three extra fields of the expansion (3.12)
are found in (3.13). However, the extra
field $b(\tau)$ is {\it not} determined by the condition (3.13). The 
equation of motion $b (\tau) + \dot e (\tau) = 0$, emerging from the Lagrangian
(2.2), comes to our help (i.e. $b = - \dot e$). With the above
insertions, the expansions in  (3.12) become
$$
\begin{array}{lcl}
&& (E)|_{(ac)}\; (\tau,  \theta) = e (\tau) 
+ \theta\; (\tilde s_{ab} e (\tau)), \qquad\;\;\;
(\Phi)|_{(ac)}\; (\tau,  \theta) = c (\tau) 
+ \; \theta\;  (\tilde s_{ab} c (\tau)), \nonumber\\
&&(\bar \Phi)|_{(ac)}\; (\tau, \theta) = \bar c (\tau) 
+ \; \theta \;(\tilde s_{ab} \bar c (\tau)). 
\end{array} \eqno(3.15)
$$
The above expansions, together with the help of the general
transformations (2.7), provide the
geometrical interpretation for the conserved and
on-shell nilpotent anti-BRST charge $\tilde Q_{ab}$ as 
the generator of translation
(i.e. $(\partial/\partial \theta)$) for the {\it anti-chiral} superfields 
(3.12) along the $\theta$-direction of the
two $(1 + 1)$-dimensional anti-chiral super sub-manifold, 
parametrized by an even
variable $\tau$ and an odd variable $\theta$. \\

\noindent
{\bf 4 Symmetries for the target space variables in superfield formalism}\\

\noindent
In contrast to the horizontality condition that relies heavily on the
(super-)exterior derivatives $(\tilde d) d$
and the (super) one-forms
$(\tilde V)A$ for the derivation of the (anti-)BRST 
symmetry transformations  on the gauge field $e(\tau)$
and the (anti-)ghost fields $(\bar c)c$, the 
corresponding nilpotent symmetries
for the target fields $(x^\mu (\tau), p_\mu (\tau))$ are obtained due to the
invariance of the conserved charge of the theory. To justify this
assertion, first of all, we start off with the super expansion of the
superfields $(X^\mu, P_\mu)(\tau, \theta,\bar\theta)$),
corresponding to the ordinary target variables $(x^\mu, p_\mu)(\tau)$
(that specify the Minkowski cotangent manifold), as
$$
\begin{array}{lcl} 
X_\mu (\tau, \theta, \bar\theta) &=& x_\mu (\tau)
\;+\; i \;\theta\; \bar R_\mu (\tau)\; +\; i \;\bar \theta \; R_\mu (\tau) 
\;+\; i \;\theta \;\bar \theta \; S_\mu (\tau),
\nonumber\\
P_\mu (\tau, \theta, \bar\theta) &=& p_\mu (\tau)
\;+\; i\; \theta \;\bar F_\mu (\tau) \;+\; i \;\bar \theta \; F_\mu (\tau) 
\;+\; i\; \theta \;\bar \theta \; T_\mu (\tau).
\end{array} \eqno(4.1)
$$
It is evident that, in the limit 
$(\theta, \bar\theta) \rightarrow 0$,
we get back the canonically conjugate
target space variables $(x^\mu (\tau), p_\mu (\tau))$ of the 
first-order Lagrangian in (2.1). Furthermore, the number of
bosonic fields ($x_\mu, p_\mu, S_\mu, T_\mu)$ do match with the fermionic
fields $(F_\mu, \bar F_\mu, R_\mu, \bar R_\mu)$ 
so that the above expansion is consistent
with the basic tenets of supersymmetry. All the component fields
on the r.h.s. of the expansion (4.1) are functions of the
monotonically increasing parameter $\tau$ of the world-line. As emphasized in
Section 2, two most decisive features of the free relativistic particle
are (i) $ \dot p_\mu = 0$, and (ii) $ p^2 - m^2 = 0$. In fact, it can be seen
that the conserved {\it gauge} charge $Q_{g} = \frac{1}{2} (p^2 - m^2)$
couples to the `gauge' (einbein) field $e(\tau)$ in the Lagrangian
density $L_f$ to maintain the local gauge invariance 
\footnote{Exactly the same kind of coupling exists for the {\it interacting}
1-form (non-)Abelian gauge theories where the matter conserved current
$J_\mu = \bar\psi \gamma_\mu \psi$, constructed by the Dirac fields,
couples to the gauge field $A_\mu$ of the (non-)Abelian
gauge theories to maintain the local gauge invariance (see, e.g., [1,2]).}
under the transformations $\delta_g x_\mu = \xi p_\mu, 
\delta_g p_\mu = 0, \delta_g e = \dot \xi$. For the BRST invariant
Lagrangian (2.2), the same kind of coupling exists for
the local BRST invariance to be maintained
in the theory. The invariance of the
condition $(p^2 - m^2 = 0$) on the supermanifold
$$
\begin{array}{lcl}
P_\mu (\tau,\theta, \bar\theta) P^\mu (\tau, \theta, \bar\theta) 
- m^2 = p_\mu (\tau) p^\mu (\tau) - m^2,
\end{array} \eqno(4.2)
$$
implies that
$$
\begin{array}{lcl}
F_\mu (\tau) = \bar F_\mu (\tau) = T_\mu (\tau) = 0, \;\;\mbox{and}\;\;
P_\mu (\tau, \theta, \bar\theta) = p_\mu (\tau). 
\end{array} \eqno(4.3)
$$
In other words, the
invariance of the mass-shell condition on the (super)manifolds
enforces $P_\mu (\tau, \theta, \bar\theta)$ to be independent of the
Grassmannian variables $\theta$ and $\bar\theta$.
To be consistent with our earlier interpretations for the (anti-)BRST charges
in the language of translation generators along the Grassmannian directions
$(\theta)\bar\theta$
of the supermanifold, it can be seen that the above equation can be 
re-expressed as
$$
\begin{array}{lcl}
P_\mu (\tau, \theta, \bar\theta) = p_\mu (\tau) 
+ \theta \; (s_{ab} p_\mu (\tau))
+\bar \theta \; (s_{b} p_\mu (\tau))
+ \theta \; \bar\theta\;(s_b s_{ab} p_\mu (\tau)).
\end{array}\eqno (4.4)
$$
The above equation, {\it vis-{\`a}-vis} (4.3),
makes it clear that $s_b p_\mu (\tau) = 0$
and $s_{ab} p_\mu (\tau) = 0$. One of the most important 
relations, that plays a pivotal role in the derivation of the mass-shell 
condition ($p^2 - m^2 = 0$) for the Lagrangian $L_f$, is
$\dot x_\mu (\tau) = e (\tau) p_\mu (\tau)$. A simpler way
to derive the (anti-)BRST  transformations on the target variables is to
require the invariance of this
central relation on the supermanifold as 
$$
\begin{array}{lcl}
\dot X_\mu (\tau,\theta,\bar\theta) = E (\tau, \theta, \bar\theta)
P_\mu (\tau, \theta, \bar\theta),
\end{array} \eqno(4.5)
$$
where $E (\tau,\theta, \bar\theta)$ is the expansion in (3.6) which has been
obtained after the application of the horizontality condition. Insertions
of the expansions in (4.1), (3.6) and (4.3) into (4.5) lead to the following 
explicit equation:
$$
\begin{array}{lcl}
\dot x_\mu (\tau) + \theta \dot {\bar R_\mu} (\tau)
+ \bar \theta \dot R_\mu (\tau)
+ i \theta\bar\theta \dot S_\mu (\tau)
= \bigl [\; e(\tau) + \theta \dot {\bar c} (\tau) + \bar \theta \dot c(\tau)
+ i \theta\bar\theta \dot b (\tau)\; \bigr ]\; p_\mu (\tau).
\end{array}\eqno (4.6)
$$
The equality of the coefficients of the appropriate terms
from l.h.s. and r.h.s. yields
$$
\begin{array}{lcl}
\dot x_\mu  = e\; p_\mu, \qquad \dot {\bar R_\mu} = \dot {\bar c}\; p_\mu,
\qquad \dot R_\mu = \dot c\; p_\mu, \qquad \dot S_\mu = \dot b\; p_\mu.
\end{array}\eqno (4.7)
$$
At this crucial stage,  we summon one of the most decisive
physical insights into the characteristic features of the free relativistic 
particle which states that there is no action of any kind of
force (i.e. $\dot p_\mu (\tau) = 0$) on the {\it free} motion of the particle. 
Having taken into account this decisive input, we obtain, from (4.7),
the following relations
$$
\begin{array}{lcl}
\dot {\bar R}_\mu \equiv
\partial_\tau {\bar R_\mu} = \partial_\tau (\bar c p_\mu), \qquad
\dot R_\mu \equiv
\partial_\tau R_\mu = \partial_\tau (c p_\mu), \qquad 
\dot S_\mu \equiv \partial_\tau
 S_\mu = \partial_\tau (b p_\mu),
\end{array}\eqno (4.8)
$$
which lead to
$$
\begin{array}{lcl}
{\bar R_\mu} (\tau) = \bar c\; p_\mu, \qquad
R_\mu (\tau) = c \;p_\mu, \qquad 
 S_\mu (\tau) = b \;p_\mu.
\end{array}\eqno (4.9)
$$
The insertions of these values in the expansion (4.1) lead to the derivation 
of the nilpotent (anti-)BRST transformations ($s_{(a)b}$) on the target space
co-ordinate field $x_\mu (\tau)$ as
$$
\begin{array}{lcl}
X_\mu (\tau, \theta, \bar\theta) &=& x_\mu (\tau)
+ \;\theta\; (s_{ab} x_\mu(\tau)) + \;\bar \theta \; (s_b x_\mu (\tau)) 
+ \;\theta \;\bar \theta \; (s_b s_{ab} x_\mu (\tau)).
\end{array}\eqno (4.10)
$$
In our recent papers [20-22] on interacting 1-form (non-)Abelian gauge 
theories, it has been shown that there is a beautiful
consistency and complementarity between the
horizontality condition and the requirement of the invariance of conserved
matter (super)currents on the (super)manifolds. The former restriction leads to
the derivation of nilpotent symmetries for the gauge- and (anti-)ghost fields.
The latter restriction yields such transformations for the matter fields. For
the case of the 
free relativistic particle, it can be seen that the invariance of
the gauge invariant and conserved charge on the (super)manifolds, leads to
the derivation of the transformations on the target field
variables. To corroborate
this assertion, we see that the conserved and gauge invariant charge
$Q_g = \frac{1}{2}\; (p^2 - m^2)$ is the analogue of the conserved {\it matter}
current of the 1-form interacting (non-)Abelian gauge theory. Since
the expansion for $P_\mu (x.\theta,\bar\theta)$ is trivial (cf. (4.3)), we
have to re-express the mass-shell condition 
(i.e. $ e^2 (p^2 - m^2)= \dot x_\mu \dot x^\mu - e^2  m^2$) in the language
of the superfields (3.6) and (4.1). Thus, the invariance of the conserved
(super)charges on the (super)manifolds is:
$$
\begin{array}{lcl}
\dot X_\mu (\tau,\theta,\bar\theta) \dot X^\mu (\tau,\theta,\bar\theta)
- m^2\;E(\tau,\theta,\bar\theta) E(\tau,\theta,\bar\theta) 
= \dot x_\mu (\tau) \dot x^\mu (\tau) - e^2\; m^2.
\end{array}\eqno (4.11)
$$
The equality of the appropriate terms from the l.h.s. and r.h.s. leads to 
$$
\begin{array}{lcl}
\dot x_\mu \dot {\bar R^\mu} = m^2\;e\; \dot {\bar c}, \qquad
\dot x_\mu \dot R^\mu = m^2\;e\; \dot c, \qquad
\dot x_\mu \dot S^\mu = m^2\;e\; \dot b, \qquad
\dot R_\mu \dot {\bar R^\mu} = m^2\;\dot c\; \dot {\bar c}.
\end{array}\eqno (4.12)
$$
Taking the help of the key relation $\dot x_\mu = e \;p_\mu$, we obtain
the expressions for $\dot R_\mu, \dot {\bar R}_\mu, \dot S_\mu$ exactly same
as the ones given in (4.7) {\it for the mass-shell condition $p^2 - m^2 = 0$ 
to be valid}. Exploiting the no force (i.e. $\dot p_\mu = 0$) criterion
on the free relativistic particle, we obtain the expressions for 
$R_\mu, \bar R_\mu, S_\mu$ in exactly the same form as given in (4.9). 
The insertion of these 
values in (4.1) leads to the same expansion as given in (4.10). This provides 
the geometrical interpretation for the (anti-)BRST charges as the 
{\it translational generators}. 
It should be noted that the
restrictions in (4.5) and (4.11) are intertwined. However, the latter
is more physical because it states the invariance of the
{\it mass-shell} condition {\it explicitly}.\\

\noindent
{\bf 5 Conclusions}\\

\noindent
We have derived,
in our present investigation, 
the off-shell (as well as on-shell) nilpotent (anti-)BRST
symmetry transformations for the target space variables 
$x^\mu (\tau)$ and $ p_\mu (\tau)$ as well as
the gauge (einbein) field ($e$) and the (anti-)ghost fields $(\bar c)c$
in the framework of augmented superfield approach to BRST formalism.
In particular, the target space fields are found to be exactly like 
the {\it matter}
(e.g. Dirac, complex scalar) fields of the usual 1-form interacting
(non-)Abelian gauge theories. This is because of the fact that,
as the conserved current, constructed by the matter fields, couples
to the 1-form gauge fields of the usual
interacting (non-)Abelian gauge theories
to maintain the {\it local} gauge invariance,
in a similar fashion, the conserved charge $Q_g = \frac{1}{2}\; (p^2 - m^2)$
(constructed by the target space variables $p_\mu$ or $\dot x_\mu$)
couples to the gauge (einbein) field $e(\tau)$ 
(cf. (2.1) or (2.2)) to maintain the
local gauge (or equivalently reparametrization) invariance in our
present theory. It has been a long-standing problem to
derive the nilpotent symmetries for the {\it matter} fields of an interacting
gauge theory in the superfield formulation
(except in our very recent endeavours [20-22]). 
We chose the system of free scalar relativistic particle
for our discussion primarily
for four reasons. First and foremost, this theory provides one of
the simplest {\it interacting} gauge theory where the reparametrization
invariance is also present. Second, this theory is at the heart of the
more general reparametrization invariant string theories which are
at the fore-front of the modern-day 
theoretical research in high energy physics. Third,
to check the mutual consistency and complementarity between the
horizontality condition and the invariance of the conserved charge/current
on the (super)manifolds which lead to the derivation of nilpotent
symmetries for {\it all} the fields of the interacting gauge theories 
(see, e.g., [20-22]). Finally, to corroborate the assertion that the
nilpotent symmetries for matter fields (in the present endeavour the target
fields) owe their origin to the invariance of the conserved charges/currents
on the (super)manifolds. The general approach of the geometrical superfield
formalism with the theoretical arsenal of (i) the horizontality condition,
and (ii) the invariance of the conserved charges/currents on the 
(super)manifold can be applied to its multi-pronged physical applications in
the context of reparametrization invariant theories such as
spinning relativistic particle, various versions of string theories,
brane dynamics and higher-spin gauge theories, etc. We intend
to pursue the above mentioned future directions to put our general
ideas of investigation on a firmer footing.
We hope to report on our key results 
of the above endeavours in our forthcoming publications.

\end{document}